\documentclass[11pt]{article}
\usepackage[margin=1in]{geometry}
\usepackage[T1]{fontenc}
\usepackage{amsmath,amssymb}
\usepackage{braket}   
\usepackage{booktabs} 
\usepackage{array}
\newcolumntype{L}[1]{>{\raggedright\arraybackslash}p{#1}}
\usepackage{graphicx}
\usepackage{tikz}
\usetikzlibrary{arrows.meta,calc,positioning}
\usepackage[numbers,sort&compress]{natbib}
\usepackage[colorlinks=true,linkcolor=blue,citecolor=blue,urlcolor=blue]{hyperref}

\newcommand{\ie}{i.e.\ }

\newcommand{\etal}{\textit{et al.}}

\newcommand{\Prob}{\mathrm{Pr}}
\DeclareMathOperator*{\argmax}{arg\,max}

\newcommand{\compactheading}[1]{\par\smallskip\noindent\textbf{#1}\quad}

\begin{document}


\title{Maximum Likelihood Decoding of Quantum Error Correction Codes}

\author{
Hanyan Cao$^{1}$, Ge Yan$^{2}$, Yuxuan Du$^{2,3,*}$, and Feng Pan$^{1,*}$\\[0.5em]
\small $^{1}$Science, Mathematics and Technology Cluster, Singapore University of Technology and Design,\\
\small 8 Somapah Road, 487372, Singapore\\
\small $^{2}$College of Computing and Data Science, Nanyang Technological University, Singapore, Singapore\\
\small $^{3}$School of Physical and Mathematical Sciences, Nanyang Technological University,\\
\small Singapore, Singapore\\[0.5em]
\small $^*$Authors to whom any correspondence should be addressed.\\
\small \texttt{yuxuan.du@ntu.edu.sg} and \texttt{feng\_pan@sutd.edu.sg}
}

\date{}

\maketitle

\begin{abstract}
Quantum error correction (QEC) is indispensable for realizing fault-tolerant quantum computation, yet its effectiveness hinges critically on the classical decoding algorithm that interprets noisy syndrome measurements.
Among all possible decoding strategies, maximum likelihood decoding (MLD) is provably optimal, since it identifies the logical group with largest likelihood by summing over all possible errors within logical class consistent with the observed syndrome.
Despite its optimality, MLD is computationally intractable in general (\#P-hard), motivating a rich landscape of exact and approximate algorithms.

In this topical review, we provide a unified perspective on MLD by surveying recent advances through three complementary lenses: statistical mechanics, tensor networks, and artificial intelligence.
From the statistical mechanics viewpoint, the MLD problem maps onto evaluating partition functions of disordered spin models, enabling exact solutions for certain codes and noise models as well as threshold estimation via phase-transition analysis.
From the tensor network perspective, approximate contraction of tensor networks on the code's factor graph yields decoders that closely approach MLD accuracy with polynomial computational cost.
From the artificial intelligence perspective, neural-network-based decoders, including autoregressive generative models and recurrent transformers, learn to approximate the MLD distribution from data, achieving high accuracy with the parallelism afforded by modern hardware accelerators.

We discuss the connections among these three approaches, review their application to both simulated and experimental quantum hardware, and outline open challenges including real-time decoding, scalability to large code distances, and generalization to high-rate quantum low-density parity-check codes.
\end{abstract}


\section{Introduction}
\label{sec:introduction}

\subsection{The promise and challenge of quantum error correction}
\label{sec:intro:promise}

Quantum computing promises exponential speedups for a range of scientifically and commercially important problems, including quantum simulation of many-body and chemical systems~\cite{nielsen_chuang_2010, quantumchemistry2005, vandamEndEndQuantumSimulation2024} and integer factorization or discrete logarithms that underpin widely deployed public-key cryptosystems~\cite{factor1997}.
Realizing this promise, however, demands that logical quantum information be protected from the inevitable noise that afflicts every physical implementation.
Contemporary quantum processors based on superconducting circuits, trapped ions, and neutral atoms have achieved high-fidelity control, but the physical error rates relevant to quantum error correction are still typically in the $10^{-3}$--$10^{-2}$ range per operation or syndrome-extraction cycle~\cite{googlequantumai2023Suppressing, googlenew2024}.
Although remarkable, these error rates remain many orders of magnitude above the $10^{-10}$--$10^{-15}$ per-operation failure probabilities estimated to be necessary for running practical quantum algorithms such as large-scale Shor factoring~\cite{factor1997, FTneed_2HowFactor20482021} or quantum chemistry simulations at chemical accuracy~\cite{quantumchemistry2005, vandamEndEndQuantumSimulation2024}.

Quantum error correction (QEC) bridges this gap by encoding logical quantum information redundantly across many physical qubits, so that errors can be detected and corrected without disturbing the encoded state~\cite{Shor1995, Steane1996, CalderbankShor1996, Gottesman1997}.
The central theoretical guarantee is the \emph{threshold theorem}: provided that the physical error rate lies below a code-dependent threshold $p_\mathrm{th}$, one can suppress the logical error rate to arbitrarily low levels by increasing the code distance $d$, at the cost of a polynomial overhead in the number of physical qubits~\cite{Dennis2002}.
This statement, however, carries an implicit but critical assumption that the decoder, \ie the classical algorithm that processes the measured error syndromes and decides on a corrective operation, is sufficiently accurate.

A decoder that is too slow or too inaccurate can erode the gains that increasing code distance is supposed to provide.
Relative to exact MLD, an approximate decoder preserves the optimal threshold and logical-error scaling only when its approximation errors are sufficiently rare on the syndromes that dominate the large-code limit; otherwise the achieved threshold or finite-size scaling can be reduced.
The decoder is, in a precise sense, the classical bottleneck of quantum error correction: it must simultaneously achieve high accuracy (to avoid misidentifying errors, which can introduce logical failures) and low latency (to keep pace with the rate at which syndromes are generated during computation, typically once every $\sim$1~$\mu$s for superconducting platforms~\cite{googlenew2024}).
Balancing these two requirements has emerged as one of the central challenges in the field.

Recent experimental milestones have underscored both the promise and the urgency of this challenge.
In 2023, Google Quantum AI demonstrated, for the first time, that increasing the code distance of a surface code from $d=3$ to $d=5$ leads to a measurable reduction in the logical error rate, establishing that their hardware operates below the surface-code threshold~\cite{googlequantumai2023Suppressing}.
A subsequent experiment in 2025 extended this to $d=7$ and beyond, demonstrating the exponential suppression of logical errors with increasing distance~\cite{googlenew2024}.
Crucially, these experiments revealed that the choice of decoder significantly influences the measured logical error rate and, by extension, the scientific conclusions drawn from the data.
Different decoders applied to the same experimental data can yield strikingly different assessments of hardware performance~\cite{PhysRevLett.134.190603}, highlighting the need for optimal or near-optimal decoding strategies.

\subsection{Decoding as an inference problem}
\label{sec:intro:inference}

At its core, the decoding problem is a problem of statistical inference.
During each round of error correction, a set of stabilizer measurements is performed, producing a binary string called the \emph{syndrome}.
The syndrome reveals the presence and approximate location of errors but cannot uniquely identify the physical error that occurred since many distinct physical errors produce the same syndrome.
The task of the decoder is, given the observed syndrome $\mathbf{s}$, to determine the most appropriate corrective operation.

The key subtlety is that not all errors with the same syndrome are equally harmful.
For the stabilizer codes that are the focus of this review, the codespace is defined as the simultaneous $+1$ eigenspace of a commuting set of Pauli checks, called stabilizers.
Two errors that differ only by multiplication by a stabilizer element act identically on the codespace and need not be distinguished.
What matters is the \emph{logical equivalence class} of the error, specifically which logical operator (if any) the error implements on the encoded information.
For a code encoding $k$ logical qubits, the set of Pauli errors consistent with syndrome $\mathbf{s}$ partitions into $4^k = 2^{2k}$ logical equivalence classes (cosets), each labeled by a $2k$-bit string $\boldsymbol{\ell}$ specifying the logical Pauli effect.
The optimal decoding strategy is to select the coset with the highest total probability, a strategy known as MLD:
\begin{equation}
\boldsymbol{\ell}_{\mathrm{MLD}} = \argmax_{\boldsymbol{\ell} \in \{0,1\}^{2k}} \Prob(\boldsymbol{\ell} \mid \mathbf{s}) = \argmax_{\boldsymbol{\ell}} \sum_{\substack{E \,:\, \sigma(E) = \mathbf{s},\\ \lambda(E) = \boldsymbol{\ell}}} \Prob(E),
\label{eq:mld}
\end{equation}
where $\sigma(E)$ denotes the syndrome of error $E$, $\lambda(E)$ its logical class, and the sum runs over all physical errors consistent with syndrome $\mathbf{s}$ and logical label $\boldsymbol{\ell}$.

Under a matched noise prior and a zero-one loss on the logical class, the MLD rule is Bayes optimal: for each observed syndrome, choosing the most probable logical class minimizes the conditional probability of a logical-class error.
However, the summation in Eq.~\eqref{eq:mld} involves an exponentially large number of terms, one for each physical error in the coset, making it computationally intractable in general.
Indeed, the problem of evaluating the MLD probability has been shown to be \#P-hard for general stabilizer codes~\cite{IyerPoulin2015}, placing it in the same computational complexity class as counting satisfying assignments or computing partition functions of classical spin models.

This connection to partition function computation is more than an analogy.
As first shown by Dennis, Kitaev, Landahl, and Preskill~\cite{Dennis2002}, the MLD problem for a stabilizer code under Pauli noise can be mapped exactly onto computing the partition function of a disordered classical spin model; for the toric or surface code under independent bit-flip noise, this model is a random-bond Ising model.
This mapping is the conceptual bridge connecting quantum error correction to the rich toolkit of statistical mechanics and tensor network methods, and it provides the organizational principle for this review.

Most practical decoders do not compute these coset probabilities exactly.
Instead, they use algorithmic surrogates that trade optimality for latency: minimum-weight perfect matching and its modern implementations~\cite{Edmonds1965, Kolmogorov2009, higgott2021pymatching, higgott2023Sparse, wu2023fusion}, Union-Find~\cite{DelfosseNickerson2021}, belief propagation and ordered-statistics post-processing~\cite{MacKay2004SparseGraph, PoulinChung2008Iterative, Roffe2020, PanteleevKalachev2021}, and correlated-noise hybrids such as correlated matching, belief matching, and matching synthesis~\cite{correlatedmatching2013, believematching2023, Jones2024MatchingSynthesis}.
These methods are indispensable baselines, but they generally optimize a different objective from degenerate MLD.
MWPM, for example, is a minimum-weight error decoder rather than a maximum-probability coset decoder; it can therefore miss the entropic contribution from many equivalent errors.
This distinction has become experimentally relevant: exact or near-exact post-processing of repetition-code and surface-code data has shown that decoder suboptimality can materially affect the logical error rates inferred from the same syndrome records~\cite{googlequantumai2023Suppressing, PhysRevLett.134.190603}.
This finding vividly illustrates why understanding and approaching MLD is of both theoretical and practical importance.

\subsection{Scope and organization of this review}
\label{sec:intro:scope}

This review surveys recent advances in MLD of quantum error-correcting codes.
Broad reviews of quantum error correction often emphasize code constructions, fault-tolerant gates, or practical decoder benchmarking for specific code families~\cite{Fowler2012, BreuckmannEberhardt2021, Roffe2020}.
Our focus is narrower and more unifying: statistical mechanics, tensor networks, and artificial intelligence are treated as complementary ways of approximating the same coset-likelihood objective, so that progress in one area directly informs the others.

The remainder of this review is organized as follows.
Section~\ref{sec:preliminaries} introduces the necessary background on stabilizer codes, noise models, the formal MLD problem, and conventional baseline decoders.
Section~\ref{sec:approaches} presents the three approaches to MLD in a unified framework: Section~\ref{sec:statmech} discusses the statistical mechanics perspective, Section~\ref{sec:tn} the tensor network perspective, and Section~\ref{sec:ai} the artificial intelligence perspective.
Section~\ref{sec:applications} then discusses applications and outlook, including noise characterization, experimental benchmarking, decoding beyond memory experiments, and open challenges.
Section~\ref{sec:conclusion} concludes with a unified outlook on the field.

\section{Preliminaries}
\label{sec:preliminaries}

This section introduces the background material needed to formulate the MLD problem precisely: the stabilizer formalism for quantum codes, the hierarchy of noise models used in practice, the formal definition of MLD, and the conventional decoders that serve as performance baselines.

\subsection{Stabilizer codes}
\label{sec:prelim:stabilizer}

The stabilizer formalism~\cite{Gottesman1997, CalderbankShor1996} provides a compact and powerful framework for describing a large class of quantum error-correcting codes.
Let $\mathcal{P}_n$ denote the $n$-qubit Pauli group, consisting of all $n$-fold tensor products of the single-qubit Pauli operators $\{I, X, Y, Z\}$, together with overall phases $\{\pm 1, \pm i\}$.
A stabilizer code $\mathcal{C}$ encoding $k$ logical qubits into $n$ physical qubits is specified by an abelian subgroup $\mathcal{S} \subset \mathcal{P}_n$, called the \emph{stabilizer group}, that contains $2^{n-k}$ elements and does not contain $-I$.
The code parameters are conventionally written as $[\![n, k, d]\!]$, where $d$ is the \emph{code distance}.
The codespace $\mathcal{H}_\mathcal{C} \subseteq (\mathbb{C}^2)^{\otimes n}$ is the simultaneous $+1$ eigenspace of all stabilizer elements:
\begin{equation}
\mathcal{H}_\mathcal{C} = \{ |\psi\rangle \in (\mathbb{C}^2)^{\otimes n} : S|\psi\rangle = |\psi\rangle \;\;\forall\; S \in \mathcal{S} \}.
\end{equation}

Errors are detected by measuring the stabilizer generators.
Each generator $S_i$ yields a binary measurement outcome $s_i \in \{0,1\}$, and the collection $\mathbf{s} = (s_1, \ldots, s_{n-k})$ is the \emph{syndrome}.
A Pauli error $E$ that commutes with all stabilizers ($s_i = 0$ for all $i$) goes undetected; if it additionally acts nontrivially on the codespace, it constitutes a logical error.
The \emph{logical operators} of the code are elements of the Pauli normalizer $N(\mathcal{S}) = \{P \in \mathcal{P}_n : PS = SP\; \forall\; S \in \mathcal{S}\}$ that are not themselves in $\mathcal{S}$.
The code distance $d$ is defined as the number of nontrivial components in the shortest logical operator; a code with distance $d$ can detect any error of weight up to $d-1$ and correct any error of weight up to $\lfloor (d-1)/2 \rfloor$.

Several families of stabilizer codes are central to this review:

\compactheading{Repetition code}
The simplest quantum error-correcting code arranges $n$ qubits along a one-dimensional chain, with $n-1$ stabilizer generators of the form $Z_i Z_{i+1}$ (or $X_iX_{i+1}$).
Despite encoding only a single logical qubit with protection against only $X$ errors (or only $Z$ errors), the repetition code plays a fundamental role as the testbed for experimental QEC demonstrations and decoding algorithm development~\cite{Kelly_2015, googlerep, googlequantumai2023Suppressing, PhysRevLett.134.190603}.

\compactheading{Surface code}
The surface (or toric) code~\cite{Kitaev2003, Bravyi1998, Dennis2002, Fowler2012} arranges qubits on a two-dimensional square lattice with stabilizers associated to faces ($X$-type) and vertices ($Z$-type).
For a rotated $d \times d$ patch, it encodes one logical qubit using $n = d^2$ data qubits (plus $d^2 - 1$ ancilla qubits for syndrome extraction) with distance $d$.
The surface code is currently the leading candidate for near-term fault-tolerant quantum computing due to its local stabilizer structure, high threshold ($p_\mathrm{th} \approx 1\%$ under circuit-level noise), and compatibility with planar qubit layouts.

\compactheading{Color code}
The color code~\cite{Bombin2006} is another two-dimensional topological code defined on a three-valent, three-colorable lattice.
Its key advantage over the surface code is the support for transversal implementations of the entire Clifford group, simplifying the overhead for fault-tolerant logical gates~\cite{Bombin2007, qec8_landahl83FaulttolerantQuantum2011, qec9_fowler2DColorCode2011}.

\compactheading{Quantum LDPC codes}
Quantum low-density parity-check (qLDPC) codes~\cite{MacKay2004SparseGraph, BreuckmannEberhardt2021} generalize topological codes by relaxing the requirement of geometric locality.
Important constructions include hypergraph-product and expander-code families~\cite{TillichZemor2014HypergraphProduct, FawziGrospellierLeverrier2018Expander}, balanced-product and quantum Tanner codes~\cite{PanteleevKalachev2022, LeverrierZemor2022}, and recent finite-size bivariate-bicycle and related group-algebra codes~\cite{PanteleevKalachev2021, qec5qccBB2024}.
These codes offer dramatically improved encoding rates compared to planar codes.
Decoding these codes is significantly more challenging, however, due to their non-planar, expander-like connectivity.

\subsection{Noise models}
\label{sec:prelim:noise}

The performance of a decoding algorithm depends critically on the assumed noise model.
In order of increasing realism, three standard noise models are used in the literature:

\compactheading{Code-capacity noise}
In the simplest setting, Pauli errors are applied independently to each data qubit, while syndrome measurements are assumed to be perfect (noiseless).
Common error channels include the \emph{bit-flip channel} (each qubit suffers an $X$ error with probability $p$), the \emph{phase-flip channel} ($Z$ errors with probability $p$), and the \emph{depolarizing channel} (each of $X$, $Y$, $Z$ occurs with probability $p/3$).
Code-capacity noise is the simplest setting for both analytical and numerical studies and is used to establish optimal thresholds; for example, the surface code under depolarizing code-capacity noise has an optimal threshold of $p_\mathrm{th} \approx 18.9\%$~\cite{Bombin_2012}.

\compactheading{Phenomenological noise}
A more realistic model supplements data qubit errors with noisy syndrome measurements: each syndrome bit may be independently flipped with some probability $q$.
Since a single unreliable syndrome measurement is insufficient, the stabilizers must be measured repeatedly over $T$ rounds, and the decoder must also infer which syndrome measurements were faulty.
The decoding problem thus acquires a temporal dimension, becoming effectively $(2+1)$-dimensional for a 2D code.
The phenomenological noise model captures the essential difficulty of noisy syndrome extraction without specifying the details of the measurement circuit.

\begin{figure}[h]
\centering
\includegraphics[width=0.9\linewidth]{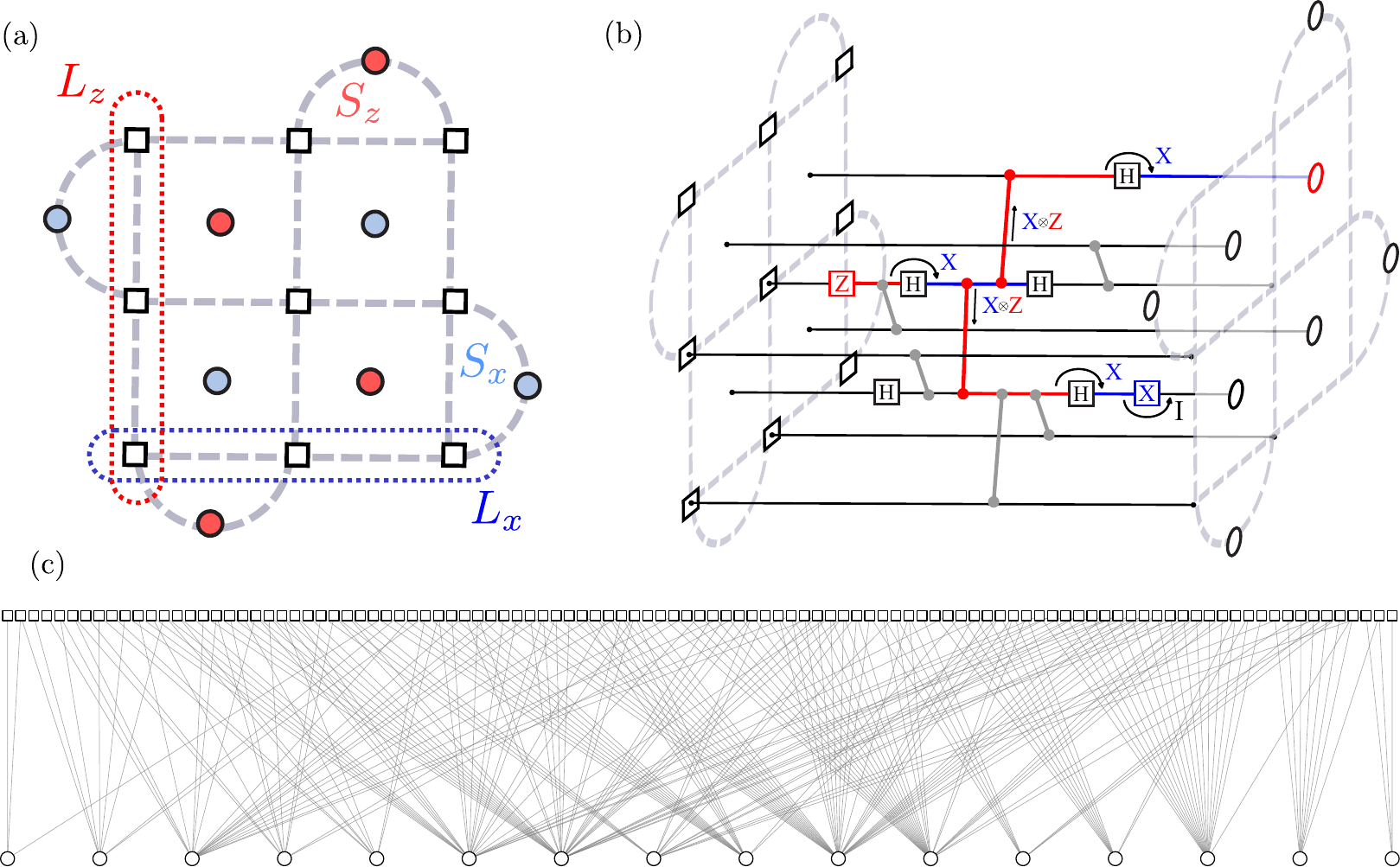}
\caption{A decoding-oriented view of the surface-code pipeline. (a) A rotated distance-3 patch with data qubits, representative $Z$-type (red) and $X$-type (blue) checks, and logical strings. (b) A local circuit fault can propagate through a syndrome-extraction circuit and create correlated detection events across space and time. (c) A DEM represents this information as a factor graph: square nodes are error mechanisms and circular nodes are detector parities.}
\label{fig:qec_cycle}
\end{figure}

\compactheading{Circuit-level noise}
The most realistic model assigns errors to every operation in the syndrome measurement circuit: state preparation, single- and two-qubit gates, idle locations, measurements, and resets. Crucially, rather than remaining isolated, these localized errors propagate forward in time through subsequent operations, spreading both spatially across the qubit array and temporally across multiple measurement rounds as shown in Fig.~\ref{fig:qec_cycle}(b).
In addition, the correlations introduced by noisy entangling gates (which can spread errors from ancilla to data qubits or between data qubits) make circuit-level noise substantially harder to decode than either code-capacity or phenomenological noise.
A widely used superconducting-inspired benchmark is the SI1000 noise-model family~\cite{mcewen2023Relaxing}, which assigns operation-dependent Pauli noise rates chosen to approximate superconducting-hardware error budgets.
Modern stabilizer simulation tools such as Stim~\cite{gidney2021Stim} facilitate the efficient analysis of circuit-level noise by extracting a \emph{detector error model} (DEM)~\cite{mcewen2023Relaxing}. The DEM is constructed by tracking the propagation of individual Pauli faults through the quantum circuit to determine which deterministic parity checks (detectors) and logical observables they flip. By compiling these physical error channels into a set of independent error mechanisms with associated probabilities, the DEM abstracts the complex spatio-temporal dynamics of circuit-level noise into a static (hyper)graphical structure suitable for decoding as shown in Fig.~\ref{fig:qec_cycle}(c).

Real devices also exhibit effects that go beyond simple Pauli DEMs, including leakage, crosstalk, quasiparticle or high-energy impact events, and slowly drifting calibration parameters~\cite{leakage2021, leakage2023, crosstalk, HighenergyImpact2024, cosmicray2024}.
This is why modern decoding papers increasingly treat noise modeling, decoder priors, and hardware calibration as a coupled problem rather than as independent stages~\cite{learning2024, cao2026differentiablemaximumlikelihoodnoise}.

\subsection{The maximum likelihood decoding problem}
\label{sec:prelim:mld}

We now define the MLD problem more precisely than in Section~\ref{sec:intro:inference}, introducing notation that will be used throughout the review.

Consider an $[\![n,k,d]\!]$ stabilizer code with stabilizer group $\mathcal{S}$.
After an error $E \in \mathcal{P}_n$ occurs, measuring the stabilizer generators produces the syndrome $\mathbf{s} = \sigma(E) \in \{0,1\}^{n-k}$.
The error also has a \emph{logical effect} $\boldsymbol{\ell} = \lambda(E) \in \{0,1\}^{2k}$, determined by its commutation relations with a chosen set of $2k$ independent logical Pauli operators.
Two errors $E_1$ and $E_2$ with the same syndrome and the same logical effect differ only by a stabilizer: $E_1 E_2^\dagger \in \mathcal{S}$.
The set of all errors consistent with syndrome $\mathbf{s}$ therefore decomposes into $2^{2k}$ \emph{cosets},
$\{C_{\boldsymbol{\ell}}(\mathbf{s})\}_{\boldsymbol{\ell} \in \{0,1\}^{2k}}$,
each labeled by a logical effect $\boldsymbol{\ell}$.
In memory experiments one often measures only one logical basis, so the reported decoding target may be a projected logical label in $\{0,1\}^{k}$ rather than the full Pauli label in $\{0,1\}^{2k}$.
This is a task-dependent readout convention, not a change in the underlying Pauli coset decomposition.

Given the noise model $\Prob(E)$, the total probability of each coset is
\begin{equation}
\Prob(\boldsymbol{\ell} \mid \mathbf{s}) = \frac{1}{\Prob(\mathbf{s})} \sum_{E \in C_{\boldsymbol{\ell}}(\mathbf{s})} \Prob(E) \;\propto\; Z_{\boldsymbol{\ell}}(\mathbf{s}),
\label{eq:coset_prob}
\end{equation}
where we have defined the \emph{coset partition function} $Z_{\boldsymbol{\ell}}(\mathbf{s}) \equiv \sum_{E \in C_{\boldsymbol{\ell}}(\mathbf{s})} \Prob(E)$.
MLD selects the coset with the highest total probability:
\begin{equation}
\hat{\boldsymbol{\ell}}_\mathrm{MLD} = \argmax_{\boldsymbol{\ell} \in \{0,1\}^{2k}} Z_{\boldsymbol{\ell}}(\mathbf{s}).
\label{eq:mld_formal}
\end{equation}

Crucially, MLD accounts for the \emph{degeneracy} of errors: even if no single error in coset $\boldsymbol{\ell}$ is particularly likely, the combined probability of all errors in that coset may exceed that of any other coset.
This is the fundamental distinction between MLD and minimum-weight decoding, which selects the single most probable error without summing over the entire coset.

The partition function $Z_{\boldsymbol{\ell}}(\mathbf{s})$ is, in general, a sum over an exponentially large number of terms: all Pauli errors that produce syndrome $\mathbf{s}$ and logical effect $\boldsymbol{\ell}$.
Computing it exactly is \#P-hard for arbitrary stabilizer codes~\cite{IyerPoulin2015}, placing MLD in the same complexity class as evaluating partition functions of general classical spin models.
However, when the code possesses special structure (planarity, bounded treewidth, or translational invariance) the partition function can often be computed exactly or approximated efficiently, which forms the subject of the remaining sections of this review.

\subsection{Conventional decoders as baselines}
\label{sec:prelim:baselines}

Before surveying MLD algorithms, we briefly review the conventional decoders that serve as performance baselines.

\compactheading{Minimum-weight perfect matching (MWPM)}
The MWPM decoder~\cite{dennis2002Topological, Edmonds1965, Kolmogorov2009, higgott2021pymatching, higgott2023Sparse, wu2023fusion} maps the decoding problem to finding a minimum-weight perfect matching on a graph whose vertices correspond to nontrivial syndrome bits and whose edge weights reflect error probabilities.
For CSS codes with independent $X$ and $Z$ noise, the two Pauli components decouple: $X$ errors are detected by $Z$-type checks and $Z$ errors are detected by $X$-type checks, so each component can be represented as a graph in which detection events are paired by candidate error chains.
The original Edmonds algorithm runs in $O(n^3)$ time; recent implementations such as PyMatching~\cite{higgott2021pymatching}, Sparse Blossom~\cite{higgott2023Sparse}, and Fusion Blossom~\cite{wu2023fusion} substantially improve practical throughput and support real-time-oriented streaming or sparse-graph decoding.
MWPM is fast and effective but does not account for degeneracy: it finds the single most probable error rather than the most probable coset, making it strictly suboptimal relative to MLD.
Its direct applicability is also limited to decoding graphs with matching structure; color codes and many qLDPC codes require transformations, projections, or different baselines.

\compactheading{Union-Find decoder}
The Union-Find decoder~\cite{DelfosseNickerson2021} achieves almost-linear time complexity $O(n\,\alpha(n))$ (where $\alpha$ is the inverse Ackermann function) by using a greedy, cluster-growth strategy.
It is one of the fastest known decoders and has been proposed as a candidate for real-time decoding in superconducting platforms.
However, its accuracy is generally lower than that of MWPM, and it shares the limitation of not accounting for degeneracy.

\compactheading{Belief propagation with ordered statistics decoding (BP+OSD)}
Belief propagation (BP) is a message-passing algorithm on the code's Tanner graph, the bipartite graph whose variable nodes represent qubits or error mechanisms and whose check nodes represent stabilizer or detector parity constraints.
It iteratively estimates the marginal error probability for each qubit~\cite{MacKay2004SparseGraph, PoulinChung2008Iterative, Roffe2020, PanteleevKalachev2021}.
Unlike MWPM, BP is applicable to arbitrary quantum codes, including qLDPC codes with non-local connectivity.
However, BP on its own struggles with the short cycles prevalent in quantum Tanner graphs and with the inherent degeneracy of quantum codes~\cite{Kuo2024}.
Ordered statistics decoding (OSD) provides a post-processing step that searches for low-weight solutions in the codespace of the reduced parity-check matrix, substantially improving accuracy~\cite{Roffe2020}.
The combined BP+OSD decoder is currently the standard baseline for qLDPC codes, though it falls short of MLD accuracy.

\compactheading{Other important baseline families}
Several additional decoders are essential for interpreting MLD results even when they are not themselves MLD solvers.
Renormalization-group and hierarchical decoders provide fast approximate decoding for topological codes~\cite{DuclosCianci2010, ducloscianci2010renormalization}.
Projection methods reduce color-code decoding to surface-code-like subproblems~\cite{Delfosse2014ColorProjection}.
Small-set-flip and related expander-code decoders give provable performance guarantees for some qLDPC families~\cite{FawziGrospellierLeverrier2018Expander}.
Belief matching, matching synthesis, and harmonized matching ensembles improve practical surface-code decoding by using richer circuit-level noise information or combining multiple approximate decoders~\cite{believematching2023, Jones2024MatchingSynthesis, Shutty2026Ensembling}.
Finally, sliding-window and parallel-window schemes address the real-time backlog problem by restructuring when and where a conventional inner decoder is applied~\cite{Skoric2023ParallelWindow, Tan2023SlidingWindow, Bombin2023Modular}.

\begin{table}[t]
\centering
\footnotesize
\setlength{\tabcolsep}{2.5pt}
\caption{Conventional decoder baselines against which MLD or near-MLD methods are commonly compared. The table emphasizes the decoding objective and the main source of suboptimality, rather than only asymptotic runtime.}
\label{tab:baseline_decoders}
\begin{tabular}{@{}L{0.16\linewidth} L{0.19\linewidth} L{0.16\linewidth} L{0.20\linewidth} L{0.17\linewidth}@{}}
\toprule
Decoder & Objective & Typical regime & Strength & Main failure mode \\
\midrule
MWPM / Sparse Blossom~\cite{Edmonds1965, higgott2023Sparse} & Minimum-weight error chain & Surface and repetition codes with matching-like DEMs & Mature, very fast, and compatible with real-time implementations & Does not sum over degenerate cosets; awkward for hyperedges and strong correlations \\
Union-Find~\cite{DelfosseNickerson2021} & Greedy cluster correction & CSS topological and erasure-like settings & Almost-linear runtime and simple local operations & Lower accuracy than matching; not likelihood calibrated \\
BP+OSD~\cite{Roffe2020, PanteleevKalachev2021} & Parity-check beliefs plus low-weight post-processing & qLDPC and nonlocal Tanner graphs & Broad applicability beyond planar matching graphs & Short cycles and degeneracy degrade BP; OSD cost can grow \\
Correlated matching / ensembles~\cite{believematching2023, Jones2024MatchingSynthesis, Shutty2026Ensembling} & Enriched matching likelihoods or decoder voting & Circuit-level surface-code data & Uses richer noise priors while retaining matching-like speed & Still approximate; ambiguous shots may require extra processing \\
\bottomrule
\end{tabular}
\end{table}

Table~\ref{tab:baseline_decoders} summarizes why these decoders remain indispensable baselines even though they do not generally implement degenerate MLD.
The MLD algorithms discussed in the following sections can be viewed as attempts to recover the missing coset-likelihood information while keeping the runtime compatible with large-scale QEC.

\section{Approaches to maximum likelihood decoding}
\label{sec:approaches}

The computational intractability of MLD for general codes has motivated a rich ecosystem of exact and approximate algorithms.
In this section, we present three major families of approaches, statistical mechanics, tensor networks, and artificial intelligence, as complementary routes to the same goal: evaluating or approximating the coset partition functions $Z_{\boldsymbol{\ell}}(\mathbf{s})$ defined in Eq.~\eqref{eq:mld_formal}.
While each approach brings its own mathematical and computational toolkit, a unifying theme is the reformulation of the discrete summation over exponentially many error configurations into a structured computation that can be performed or approximated efficiently.

\begin{figure}[h]
\centering
\resizebox{0.98\linewidth}{!}{%
\begin{tikzpicture}[
  node distance=1.0cm,
  roadbox/.style={draw, rounded corners=2pt, align=center, inner sep=5pt, font=\small, text width=4.2cm, minimum height=1.25cm},
  centerbox/.style={draw, rounded corners=2pt, align=center, inner sep=6pt, font=\small, text width=5.1cm, minimum height=1.3cm, fill=gray!8},
  arrow/.style={-{Latex[length=2.2mm]}, thick}
]
\node[centerbox, fill=gray!10] (input) at (0,2.7) {\textbf{Input}\\ syndrome $\mathbf{s}$, code, and noise prior $\Prob(E)$};
\node[centerbox, fill=yellow!12] (mld) at (0,0.8) {\textbf{MLD objective}\\ compute or approximate $Z_{\boldsymbol{\ell}}(\mathbf{s})$ and choose $\argmax_{\boldsymbol{\ell}} Z_{\boldsymbol{\ell}}$};
\node[roadbox, fill=blue!8] (stat) at (-5.2,-1.4) {\textbf{Statistical mechanics}\\ spin-model mapping, partition functions, thresholds};
\node[roadbox, fill=green!8] (tn) at (0,-1.4) {\textbf{Tensor networks}\\ factor graph, controlled contraction, bond dimension $\chi$};
\node[roadbox, fill=red!7] (ai) at (5.2,-1.4) {\textbf{Artificial intelligence}\\ learned posterior, streaming decision rules, hardware parallelism};
\node[centerbox, fill=gray!10] (output) at (0,-3.5) {\textbf{Outputs}\\ logical decision, threshold benchmark, or calibrated decoder};
\draw[arrow] (input) -- (mld);
\draw[arrow] (mld) -- (stat);
\draw[arrow] (mld) -- (tn);
\draw[arrow] (mld) -- (ai);
\draw[arrow] (stat) -- (output);
\draw[arrow] (tn) -- (output);
\draw[arrow] (ai) -- (output);
\end{tikzpicture}%
}
\caption{Roadmap of the three perspectives used in this review.  Each approach is organized around the same coset-likelihood object but uses a different computational representation: a spin-model partition function, a tensor-network contraction, or a learned posterior/decision rule.}
\label{fig:roadmap}
\end{figure}
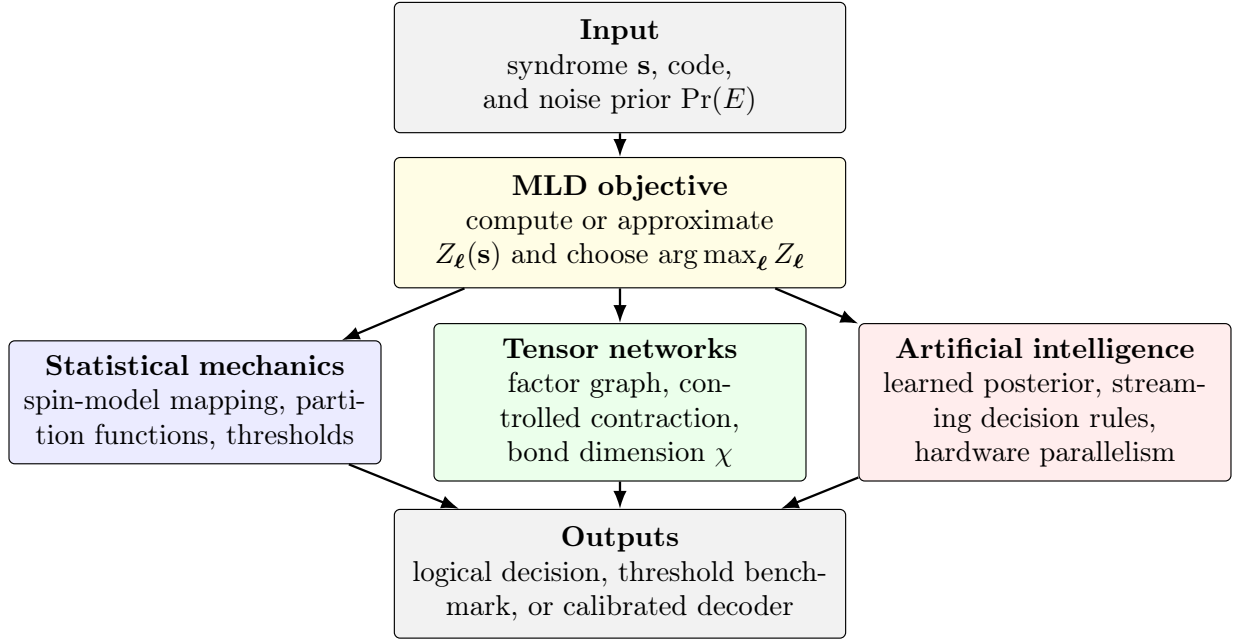

\subsection{Statistical mechanics perspective}
\label{sec:statmech}

This subsection describes how the MLD problem maps onto computing partition functions of disordered classical spin models, how this mapping enables exact solutions for certain code families, and how it provides tools for both approximate decoding and threshold estimation.

\subsubsection{Mapping QEC to statistical mechanics}
\label{sec:statmech:mapping}

The foundational observation, due to Dennis, Kitaev, Landahl, and Preskill~\cite{Dennis2002}, is that the coset partition function $Z_{\boldsymbol{\ell}}(\mathbf{s})$ defined in Eq.~\eqref{eq:coset_prob} can be rewritten as the partition function of a classical spin model defined on a graph determined by the code's structure.

Consider a stabilizer code under independent Pauli noise.
Each qubit $i$ has an error probability vector $(p_I^{(i)}, p_X^{(i)}, p_Y^{(i)}, p_Z^{(i)})$, and the total error probability factorizes as $\Prob(E) = \prod_i p_{E_i}^{(i)}$.
Taking the logarithm turns this product into a sum of local negative log-likelihoods, so, up to a normalization constant, the same probability can be written as a Boltzmann weight $\Prob(E) \propto e^{-\beta H(E)}$.
In this representation $H(E)$ is the Hamiltonian of a random-bond Ising-type model and $\beta$ is an inverse-temperature convention determined by the error rates.
The bonds (couplings) of the spin model are ``random'' in the sense that their signs are determined by a representative error compatible with the observed syndrome $\mathbf{s}$: nontrivial syndrome bits enforce frustration in the corresponding interaction pattern.

The coset partition function then takes the form
\begin{equation}
Z(\boldsymbol{\ell},\mathbf{s}) = \sum_{\{\sigma\}} e^{-\beta H_{\boldsymbol{\ell},\mathbf{s}}(\{\sigma\})},
\label{eq:statmech_partition}
\end{equation}
where the sum runs over all spin configurations corresponding to stabilizer generators, and $H_{\boldsymbol{\ell},\mathbf{s}}$ is the disorder-dependent Hamiltonian.
MLD is therefore equivalent to determining which sector has the larger partition function.
\begin{figure}[ht]
\centering
\includegraphics[width=0.8\linewidth]{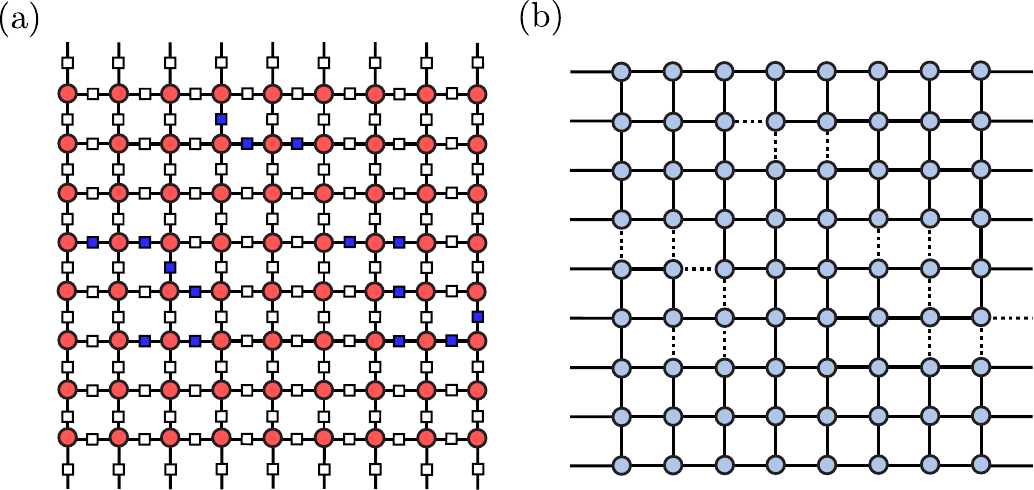}
\caption{The statistical-mechanical mapping for quantum error correction.
(a) $d=9$ surface code under bit flip code capacity noise. Each red node is a $Z$ type stabilizer, which detects $X$ error on neighbor qubits (squares). The blue squares represent that $X$ errors occur on corresponding qubits. 
(b) The dual spin-glass model corresponding to (a). Each blue node represents an $X$ spin, where a spin state of 1 indicates the application of the corresponding $X$ stabilizer. And the dashed lines are negative couplings corresponding to blue squares in (a).}
\label{fig:statmech_mapping}
\end{figure}

To make this mapping concrete, consider the toric code under independent bit-flip noise with error probability $p$ as illustrated in Fig.~\ref{fig:statmech_mapping}.
Each data qubit sits on an edge of a square lattice, and the $Z$-type stabilizers correspond to the vertices.
An $X$ error on qubit $i$ anticommutes with exactly the two vertex stabilizers adjacent to edge $i$.
The probability of an error configuration $\mathbf{e} = (e_1, \ldots, e_n) \in \{0,1\}^n$ (where $e_i = 1$ indicates an $X$ error on qubit $i$) is $\Prob(\mathbf{e}) = \prod_i p^{e_i}(1-p)^{1-e_i}$.
Now introduce Ising spins $\tau_f = \pm 1$ on each face $f$ of the lattice (the dual lattice sites).
Each error configuration can be parameterized by the domain walls of the spin configuration: $e_i = 1$ if and only if the two face-spins adjacent to edge $i$ disagree ($\tau_f \tau_{f'} = -1$).
Substituting this parameterization into the error probability yields
\begin{equation}
\Prob(\mathbf{e}[\tau]) \propto \prod_{\langle f, f' \rangle} \exp\!\big(K\, \eta_{ff'}\, \tau_f \tau_{f'}\big),
\label{eq:rbim}
\end{equation}
where $K = \frac{1}{2}\ln\!\big(\frac{1-p}{p}\big)$ is the coupling constant, $\langle f,f' \rangle$ runs over nearest-neighbor face pairs sharing edge $i$, and $\eta_{ff'} = \pm 1$ encodes the $X$ error: $\eta_{ff'} = -1$ if the $X$ error occurs on the qubit between faces $f$ and $f'$, and $\eta_{ff'} = +1$ otherwise.
Equation~\eqref{eq:rbim} is the random-bond Ising model (RBIM) form of the decoding problem.
Under this statistical mapping, the probability of a specific Pauli error configuration is directly connected to the energy of the corresponding RBIM evaluated at the uniform ferromagnetic state (where $\tau_f = +1$ for all $f$).
Crucially, multiplying the error by elements of the stabilizer group corresponds exactly to flipping the spins $\tau_f$ in the dual lattice.
Consequently, the total probability of an error equivalence class (or logical coset) is strictly proportional to the full partition function of this RBIM.
This equivalence allows established theoretical results from statistical mechanics to be imported into QEC.
Most notably, the theory of phase transitions provides a framework for analyzing code performance: the optimal error threshold of the quantum code maps to the thermodynamic phase transition separating the ordered ferromagnetic phase (where decoding succeeds) from the disordered paramagnetic phase (where decoding fails) in the RBIM.

\subsubsection{The Nishimori line and error thresholds}
\label{sec:statmech:nishimori}

A crucial feature of this mapping is that the physical error rate uniquely determines both the coupling strengths and the temperature of the spin model.
The resulting parameter line in the disorder--temperature phase diagram is the \emph{Nishimori line}~\cite{Nishimori1980}, along which the system enjoys special symmetry properties.

The Nishimori line is defined by the condition $\beta = \beta_N$, where $\beta_N$ is the inverse temperature that generated the quenched disorder. In the QEC context, this means that the decoder's assumed noise model perfectly matches the actual noise.
Along this line, several remarkable identities hold: (i)~the quenched and annealed averages of certain observables coincide, (ii)~the internal energy is self-averaging, and (iii)~the probability of a specific domain-wall pattern equals the Boltzmann weight of the corresponding spin configuration.
For QEC, property~(iii) has a direct physical interpretation: when the decoder perfectly knows the noise model, the probability that a specific error pattern occurred is exactly the statistical weight the decoder assigns to it.
This is precisely the condition under which MLD is well-calibrated.

The error threshold of the code corresponds to the point at which the Nishimori line crosses through a phase transition of the disordered spin model:

\begin{itemize}
  \item \textbf{Below threshold} ($p < p_\mathrm{th}$): the system is in an ordered phase. The partition function is dominated by the correct sector, and MLD succeeds with high probability.
  \item \textbf{Above threshold} ($p > p_\mathrm{th}$): the system enters a disordered phase. Errors proliferate and no decoder (including MLD) can reliably correct them in the asymptotic limit.
\end{itemize}

This identification of error thresholds with phase transitions along the Nishimori line has been one of the most powerful insights connecting QEC to condensed matter physics.
It allows the vast machinery of statistical mechanics, Monte Carlo simulations, transfer matrix methods, series expansions, duality arguments, to be brought to bear on the problem of estimating optimal thresholds.
Dennis \etal~\cite{Dennis2002} used this approach to establish the optimal threshold for the toric code under independent bit-flip noise at $p_\mathrm{th} \approx 10.9\%$, matching the critical point of the RBIM on the square lattice along the Nishimori line.
This remains one of the most precisely known thresholds in quantum error correction.


Chubb and Flammia~\cite{ChubbFlammia2021} significantly generalized the Dennis--Kitaev--Landahl--Preskill mapping to arbitrary stabilizer and subsystem codes under \emph{correlated} Pauli noise, including noise models arising from fault-tolerant syndrome extraction circuits.
Their formulation shows that correlations in the noise model translate into multi-body interactions in the spin model, and that the resulting statistical-mechanical models can still be analyzed using standard techniques from the theory of disordered systems.
Importantly, their work also makes explicit the connection between the statistical-mechanical partition function and tensor-network-based decoding algorithms (discussed in Section~\ref{sec:tn}), providing a unified mathematical framework that bridges the two perspectives.
For the surface code, they demonstrated that ``bunching'' correlations (where nearby qubits are more likely to err together) reduce the threshold from approximately $10.9\%$ under independent noise to approximately $10.0\%$, while ``spreading'' correlations can in some cases improve the threshold~\cite{Tuckett2020}.
These quantitative predictions are directly testable on experimental hardware and provide important guidance for hardware design.

Table~\ref{tab:thresholds} collects representative optimal thresholds obtained via statistical-mechanical mappings.
These values should not be read as universal constants of the code alone: the threshold depends on the error parametrization, measurement model, correlations, and decoder objective.
Within a fixed model, however, the optimal threshold is a fundamental asymptotic benchmark, so comparing the achieved threshold of a practical decoder against these values quantifies the room for improvement.

\begin{table}[t]
\centering
\small
\setlength{\tabcolsep}{3pt}
\caption{Representative optimal thresholds and the assumptions under which they apply.  The values indicate the maximum error rate below which MLD can make the logical error rate vanish in the large-distance limit for the specified noise model.}
\label{tab:thresholds}
\begin{tabular}{@{}L{0.18\linewidth} L{0.21\linewidth} L{0.27\linewidth} c L{0.18\linewidth}@{}}
\toprule
Code & Noise setting & Key assumptions & $p_\mathrm{th}$ & Method / ref. \\
\midrule
Toric / surface & Bit-flip code-capacity & Independent $X$ errors and perfect syndrome measurements & $10.9\%$ & RBIM mapping~\cite{Dennis2002} \\
Toric / surface & Depolarizing code-capacity & Independent $X,Y,Z$ data errors and full Pauli cosets & $18.9\%$ & Duality and numerics~\cite{Bombin_2012} \\
Toric / surface & Phenomenological & Data errors plus noisy repeated syndrome measurements & $3.3\%$ & 3D random-plaquette mapping~\cite{Dennis2002} \\
Surface code & Correlated Pauli code-capacity & Local bunching correlations in the disorder model & ${\sim}10.0\%$ & Correlated-noise mapping~\cite{ChubbFlammia2021} \\
\bottomrule
\end{tabular}
\end{table}

Beyond these canonical thresholds, the same statistical-mechanical machinery has been used to study erasures, biased noise, and code deformations tailored to asymmetric channels~\cite{Ohzeki2012, tuckett2019tailoring, Tuckett2020, qec4XZZXSurface2021}.
These works are important for a decoding review because they show that ``the'' threshold is not a property of the code alone; it depends on the noise bias, measurement circuit, and decoder objective.

\subsubsection{Exact solutions on planar graphs}
\label{sec:statmech:exact}
While the evaluation of the partition function in Eq.~\eqref{eq:statmech_partition} is \#P-hard in general, a remarkable exception arises when the underlying graph is \emph{planar}. A classical result in statistical mechanics, initially formulated by Kac and Ward~\cite{kac1952combinatorial}, establishes that the partition function of an Ising model on a planar graph can be computed exactly in polynomial time without resorting to intractable combinatorial sums.

The Kac--Ward formulation achieves this by mapping the high-temperature loop expansion of the Ising partition function directly to the determinant of a specialized linear operator. Given an Ising model on a planar graph $G = (V, E)$ with local coupling constants $\{J_{ij}\}$, one constructs the \emph{Kac--Ward matrix} $K$. Unlike standard adjacency matrices, $K$ is defined on the $2|E|$ directed edges of the graph. Its non-zero entries encode both the local edge weights $\tanh(J_{ij})$ and a geometric phase factor $\exp(i \Delta \theta / 2)$, which meticulously tracks the turning angle $\Delta \theta$ between contiguous directed edges in the planar embedding. This phase assignment correctly enumerates the parity of self-intersecting loops. The partition function is then precisely given by $Z \propto \sqrt{\det(K)}$. Because the determinant of a $2|E| \times 2|E|$ matrix can be computed in $\mathcal{O}(|E|^3)$ time, this provides a highly efficient exact algorithm.
For exact MLD, one must evaluate these partition functions across different topological sectors, which correspond to the distinct logical equivalence classes $\ell$. In the Kac--Ward framework, this is elegantly achieved by introducing discrete branch cuts into the graph---specifically, by flipping the signs of the couplings $\tanh(J_{ij})$ along the non-contractible cycles that represent the logical operators. Consequently, for a code encoding $k$ logical qubits, the optimal MLD decoding translates into evaluating $2^{2k}$ separate Kac--Ward determinants.

This connection to matchgate simulation~\cite{Valiant2002} was exploited in the context of quantum error correction by Bravyi, Suchara, and Vargo~\cite{Bravyi2014}, who showed that surface-code decoding can be formulated in ways that are exactly solvable for special planar cases and accurately approximated by tensor-network contraction for more general Pauli noise.
Their work provided one of the first concrete demonstrations that degeneracy-aware likelihood evaluation can outperform minimum-weight decoding while remaining computationally practical.

The crucial advance by Cao \etal~\cite{PhysRevLett.134.190603} was the observation that the \emph{repetition code under circuit-level noise} also maps to a planar graph spin-glass problem, despite the three-dimensional structure of the decoding problem (data qubits $\times$ syndrome rounds).
This planarity arises because the DEM of the repetition code, even under the full SI1000 circuit-level noise model~\cite{mcewen2023Relaxing}, has a graph structure that can be embedded in the plane after appropriate edge contractions.
The key technical insight is that the hyperedges in the DEM, which represent correlated errors introduced by two-qubit gates, connect at most three detectors, and the resulting hypergraph remains planar for the one-dimensional repetition code geometry.
Their ``Planar'' algorithm computes exact MLD coset probabilities by reducing the problem to planar Ising partition functions; direct determinant-based planar-Ising solvers give polynomial worst-case complexity, cubic in the planar graph size for standard dense linear-algebra implementations.

This exact solution yielded several notable results.  It computed optimal repetition-code thresholds under circuit-level depolarizing and SI1000-inspired noise models; it showed, by re-analyzing Google repetition-code data, that matching-based post-processing can overestimate logical error rates in some regimes; it was applied to a repetition-code dataset from a 72-qubit superconducting processor without reset gates, where the more correlated DEM nevertheless retains the planar structure required by the solver; and it connects, in the independent code-capacity limit, to exactly solvable surface-code thresholds such as the $p_\mathrm{th}\approx 10.9\%$ bit-flip threshold.

This work demonstrated that exact MLD can be achieved with polynomial computational complexity for a practically relevant quantum error-correcting code under a realistic noise model, shifting the perspective on MLD from a purely theoretical benchmark to a practical decoding strategy.

\subsubsection{Approximate methods: Monte Carlo and belief propagation}
\label{sec:statmech:approximate}

When the spin-model graph is not planar, as is the case for the surface code under circuit-level noise, or for qLDPC codes, exact evaluation of the partition function is intractable, and one must resort to approximate methods.

\compactheading{Monte Carlo sampling}
Markov chain Monte Carlo (MCMC) methods provide a natural approach to estimating partition function ratios.
Given two cosets labeled $\boldsymbol{\ell}_0$ and $\boldsymbol{\ell}_1$, one can estimate the ratio $Z(\boldsymbol{\ell}_1, \mathbf{s}) / Z(\boldsymbol{\ell}_0, \mathbf{s}) $ by sampling spin configurations from the Boltzmann distribution of one sector and computing the relative weight of the other.
In practice, this is typically implemented via simulated tempering or parallel tempering, where multiple replicas at different temperatures are simulated simultaneously to overcome free-energy barriers.
While MCMC sampling is asymptotically unbiased, its convergence can be slow near the phase transition (\ie near the error threshold), precisely the regime of most practical interest.
The autocorrelation time $\tau_\mathrm{auto}$ grows as a power law in the system size near criticality, $\tau_\mathrm{auto} \sim L^z$ with dynamical exponent $z \approx 2$--$3$, making the approach computationally expensive for large code distances.
Despite these limitations, Monte Carlo, transfer-matrix, and duality-based calculations serve as important benchmarks for validating other approximate MLD algorithms, especially in regimes where no exact solution is available~\cite{Bombin_2012, Ohzeki2012, ChubbFlammia2021}.

\compactheading{The BlockBP decoder}
A particularly promising approximate approach is the BlockBP decoder, introduced by Kaufmann and Arad~\cite{KaufmannArad2025BlockBP}.
The key idea is to replace the expensive matrix product state (MPS) contraction used in tensor network decoders (see Section~\ref{sec:tn}) with the BlockBP algorithm, an approximate tensor network contraction scheme based on belief propagation on a block graph.

The construction proceeds in two steps.
First, the tensor network representing the coset partition function is partitioned into \emph{blocks}, contiguous subsets of tensors that are contracted exactly within each block, yielding a coarser ``block graph.''
Second, standard belief propagation is run on this block graph, where the messages are now tensors (or low-rank approximations thereof) rather than scalar probabilities.
This two-level structure exploits the local accuracy of exact contraction within blocks while using BP to propagate information globally across the lattice.

The decoder operates within the framework of \emph{degenerate} MLD, \ie it computes coset probabilities rather than seeking a minimum-weight error, but uses the computationally cheaper BP message-passing instead of exact MPS contraction.
The advantages of BlockBP over conventional tensor network decoders are twofold.
First, the BP updates are inherently parallel, making the algorithm well-suited for GPU or FPGA acceleration and thus a candidate for real-time decoding.
Second, the algorithmic complexity scales favorably with the code size.
Numerical experiments on the surface code under depolarizing noise show that BlockBP outperforms MWPM by more than an order of magnitude in logical error rate at moderate code distances, approaching the performance of exact MLD.
The BlockBP decoder beautifully illustrates the synergy between the statistical mechanics and tensor network perspectives: it uses the graphical model structure from the stat-mech mapping together with the contraction machinery from the tensor network framework.

\subsection{Tensor network perspective}
\label{sec:tn}

A complementary perspective formulates the MLD problem directly as the contraction of a tensor network.
This approach builds on the formal equivalence between QEC decoding and TN contraction identified by Ferris and Poulin~\cite{FerrisPoulin2014}, the surface-code MLD algorithms of Bravyi, Suchara, and Vargo~\cite{Bravyi2014}, and later complexity-theoretic and general-code formulations~\cite{bohdanowiczTensorNetworks2019, Chubb2021tn2d}.
It leverages machinery developed in condensed matter physics for simulating strongly correlated quantum systems.
Its advantages are systematic accuracy control via the bond dimension, natural compatibility with automatic differentiation, and a clear connection to the statistical-mechanical formulation.
\subsubsection{MLD under the generator picture}

\begin{figure}[ht]
\centering
\includegraphics[width=0.8\linewidth]{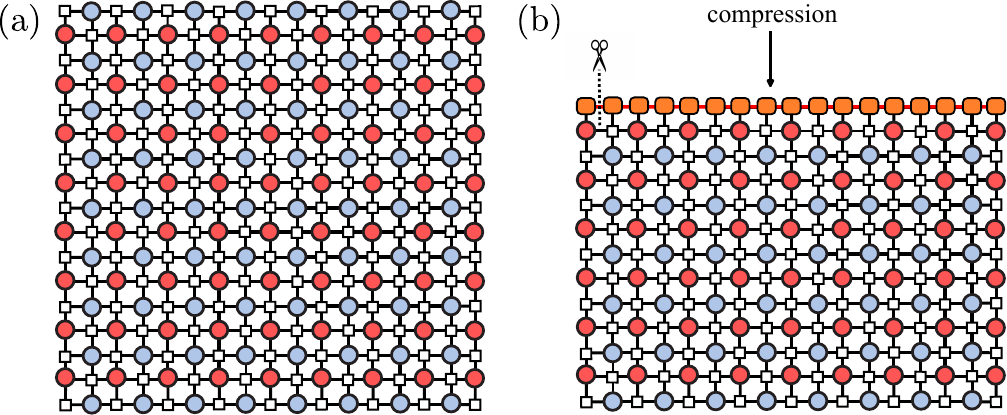}
\caption{(a) Tensor network representation of a surface code ($d=9$). (b) Boundary MPS algorithm for approximate contraction.
}
\label{fig:gtn}
\end{figure}

In the generator picture, tensor-network-based MLD relies on the fundamental degeneracy of stabilizer codes: multiplying a fixed representative error $E_{\mathbf{s},\boldsymbol{\ell}}$ by any stabilizer $S\in\mathcal{S}$ changes neither its syndrome nor its logical label. Consequently, the coset summation in Eq.~\ref{eq:coset_prob} can be rewritten as a sum over stabilizer elements. Since the stabilizer group is Abelian and generated by independent binary choices, this sum factorizes into binary variables $\alpha_i$.

\begin{equation}
\begin{split}
        Z(\boldsymbol{\ell}, \mathbf{s})  &= \sum_{S\in\mathcal{S}} \Prob(E_{\boldsymbol{\ell},\mathbf{s}}\cdot S)\\
        &=\sum_{\{\alpha_i\}} \Prob\!\left(E_{\boldsymbol{\ell},\mathbf{s}}\cdot\prod^{n-k}_{i=1} S^{\alpha_i}_i\right) \;,
\end{split}
\end{equation}
where the $S_i$ are stabilizer generators. This local sum-of-products form naturally maps onto a tensor network contraction, where the binary variables $\alpha_i$ serve as the internal indices connecting the tensors. Fig.~\ref{fig:gtn} (a) illustrates an example tensor network for a distance-9 surface code under code-capacity noise. In this diagram, each blue (red) circle corresponds to an $X$-type ($Z$-type) stabilizer and is formulated as a copy tensor, whose elements are nonzero only when all connected indices are identical.
Conversely, the white squares correspond to physical qubits and are defined as probability tensors. The elements of these probability tensors encode the probabilities of specific Pauli faults acting on the respective physical qubits~\cite{Bravyi2014}. The tensor network constructed in this manner inherently preserves the topological structure of the original quantum code. Consequently, for codes possessing a planar geometry, such as the surface code, decoding can be effectively executed using well-established tensor network contraction algorithms, such as boundary MPS contraction as shown in Fig.~\ref{fig:gtn} (b). The MPS boundary contraction algorithm proceeds as follows.
Consider a 2D tensor network arranged on an $L_x \times L_y$ grid.
The algorithm initializes the boundary MPS as the first row of the tensor network.
For each subsequent row $j = 2, \ldots, L_y$: (i)~each tensor in row $j$ is contracted with the corresponding site of the boundary MPS, creating a new MPS with enlarged bond dimension $\chi' = \chi \cdot q$, where $q$ is the local physical dimension; (ii)~the resulting MPS is compressed back to a maximum bond dimension $\chi_{\max}$ using successive singular value decomposition (SVD) truncations that discard the smallest singular values.
After all rows have been contracted, the resulting scalar (or small tensor) encodes the partition function $Z_{\boldsymbol{\ell}}(\mathbf{s})$ up to the approximation error introduced by the bond dimension truncation.
The computational cost per row is $O(L_x \cdot \chi_{\max}^3 \cdot q^2)$, yielding a total complexity of $O(L_x \cdot L_y \cdot \chi_{\max}^3 \cdot q^2)$ for the full contraction.
For binary-index surface-code constructions, $q=2$; depolarizing or correlated-noise models can increase local tensor ranks and prefactors.

Bravyi, Suchara, and Vargo~\cite{Bravyi2014} demonstrated that for the surface code under depolarizing noise, MLD can be approximated with high accuracy using MPS contraction.
They found that a relatively small bond dimension (e.g., $\chi_\mathrm{max} \approx 20$) is sufficient to achieve logical error rates indistinguishable from exact MLD, dramatically outperforming standard MWPM.
Crucially, the computational complexity of MPS contraction scales as $O(n \chi^3)$, providing a polynomial-time approximation to the \#P-hard MLD problem.

The rapid convergence with $\chi_{\max}$ can be understood through the lens of entanglement in the mapped spin model.
In the ordered phase (below the error threshold), the boundary MPS has limited entanglement, correlations decay exponentially with a finite correlation length $\xi$, and moderate $\chi_{\max}$ captures the essential physics.
Near the phase transition, the correlation length grows and ultimately diverges, requiring larger $\chi_{\max}$, but the growth remains manageable away from criticality.
Systematic benchmarks showed that at $p = 5\%$ (well below the ${\sim}18.9\%$ optimal threshold), $\chi_{\max} = 8$ suffices for distances up to $d = 25$, while at $p = 15\%$, one needs $\chi_{\max} \approx 32$--$64$ for comparable accuracy~\cite{Bravyi2014}.

This approach was later generalized by Chubb~\cite{chubb2021General} to arbitrary 2D Pauli codes under general Pauli noise models.
By using a modular tensor construction that separates the code structure from the noise model, Chubb's algorithm provides a broadly applicable, degeneracy-aware decoder for 2D stabilizer and subsystem-code settings, including color codes, Bacon--Shor-type geometries, and codes with boundaries.

\subsubsection{MLD under the detector picture}
\label{sec:tn:formulation}

On the other hand, constructing a tensor network based on the generator picture is challenging for the DEMs frequently encountered in modern experiments. This difficulty arises because DEMs provide only parity-check information, lacking the complete algebraic structure of the stabilizer group. Since MLD requires summing over degrees of freedom that leave the syndrome invariant, one must explicitly identify these summation variables. While it is possible to automatically extract these variables for simple DEM structures, such as repetition codes, doing so becomes notoriously difficult for surface codes and more complex DEM architectures. Consequently, we must approach the MLD summation problem from an alternative perspective. Whereas the generator picture builds the equivalence class by repeatedly applying stabilizers to a fixed Pauli operator, the \emph{detector picture} starts from the complete Pauli group space and progressively restricts the configuration space using syndrome constraints until all parity conditions are met.

\begin{figure}[ht]
    \centering
    \includegraphics[width=0.8\linewidth]{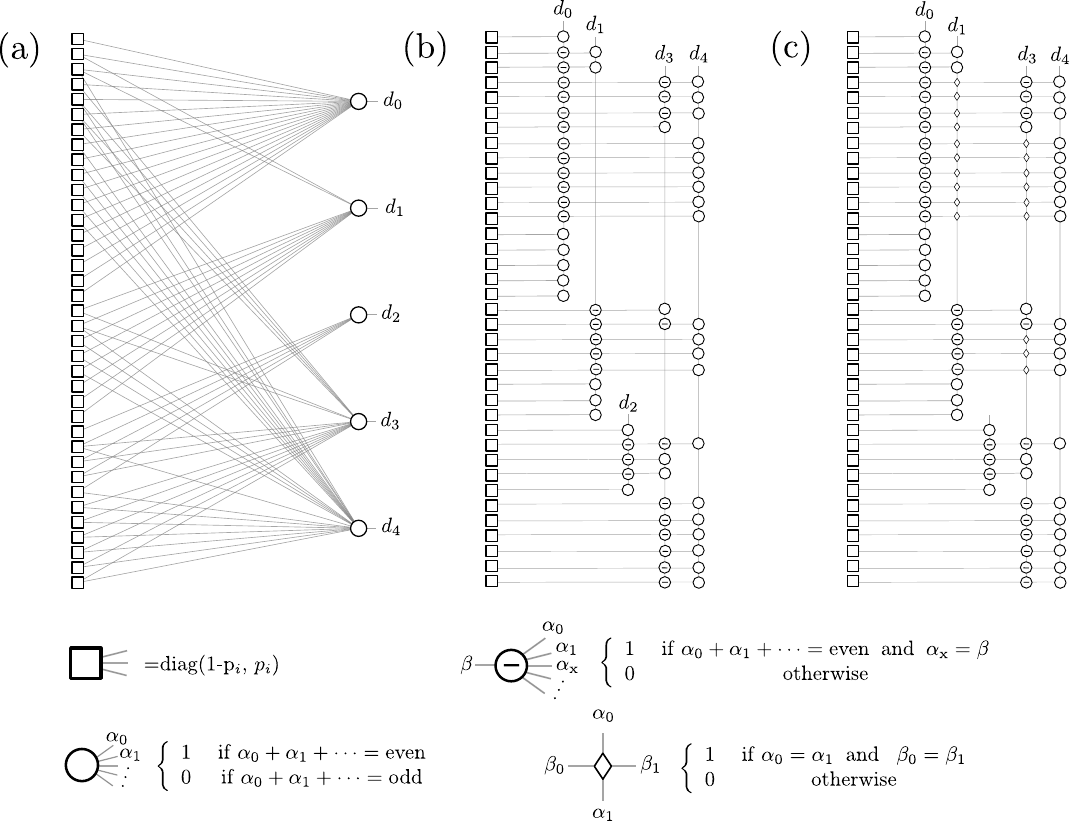}
    \caption{Tensor network in detector picture. (a) Tensor network representation of a DEM comprising 5 detectors and 37 error mechanisms. (b) Decomposition of high-degree XOR tensors into multiple lower-degree XOR tensors. (c) Insertion of passing tensors to resolve edge crossings and construct a regular 2D lattice.}
    \label{fig:dtn}
\end{figure}

Under this detector picture, the probability $\Pr(E)$ of a Pauli error configuration, together with the constraints imposed by the observed syndrome $\mathbf{s}$, can be naturally encoded in a factor graph as illustrated in Fig.~\ref{fig:dtn}(a). In this graph, square nodes represent the errors on physical qubits (or specific error mechanisms in a DEM), and circle nodes represent the stabilizers or detectors (which enforce the syndrome constraints). By assigning a tensor to each node, the sum over all error configurations consistent with the syndrome becomes the contraction of the resulting tensor network:
\begin{equation}
Z_{\boldsymbol{\ell}}(\mathbf{s}) = \text{tTr}\left( \bigotimes_i T^{(i)}_{\text{qubit}} \bigotimes_j T^{(j)}_{\text{stab}}(\mathbf{s}) \bigotimes_m T^{(m)}_{\text{logical}}(\boldsymbol{\ell}) \right),
\label{eq:tn_contraction}
\end{equation}
where $\text{tTr}$ denotes the tensor trace (contraction of all bonded indices). Here, the qubit tensors $T_{\text{qubit}}$ encode the local error probabilities, the stabilizer tensors $T_{\text{stab}}$ act as projection operators that enforce the parity bits specified by the syndrome $\mathbf{s}$, and the logical tensors $T_{\text{logical}}$ apply boundary conditions to restrict the sum to a specific logical sector $\boldsymbol{\ell}$.

Concretely, each qubit or error-mechanism tensor can be represented, in the binary independent-error case, by a diagonal matrix $T^{(i)}_{\text{qubit}} = \mathrm{diag}(1-p_i,\, p_i)$ that assigns the appropriate probability to the ``no error'' and ``error'' states. Each detector tensor $T^{(j)}_{\text{stab}}$ is a generalized Kronecker delta tensor that is nonzero only when the parity of its incident indices matches the observed syndrome bit $s_j$:
\begin{equation}
T^{(j)}_{\text{stab}}(e_{i_1}, \ldots, e_{i_w}; s_j) = \delta\!\big(e_{i_1} \oplus \cdots \oplus e_{i_w},\; s_j\big),
\end{equation}
where $w$ is the weight of stabilizer $S_j$ and the indices $i_1, \ldots, i_w$ label the qubits in its support. The logical tensors similarly enforce a parity constraint along a chosen representative of each logical operator.

Although casting the problem as a tensor network opens the door to boundary-contraction methods with tunable accuracy, direct contraction is nontrivial for realistic DEMs. The resulting networks can be dense, high-degree, and non-planar. Practical implementations therefore preprocess the network before applying an MPS-style contraction~\cite{piveteau2023Tensorbeyond2d, cao2026differentiablemaximumlikelihoodnoise}. Mapping an irregular tensor network [as depicted in Fig.~\ref{fig:dtn}(a)] onto a more regular two-dimensional layout proceeds via two representative transformations. First, as illustrated in Fig.~\ref{fig:dtn}(b), the associativity of parity constraints allows a high-degree detector tensor to be decomposed into a sequence of lower-degree XOR tensors. Second, because the resulting graph can contain crossing edges, auxiliary passing tensors can be inserted at crossings, as shown in Fig.~\ref{fig:dtn}(c). These passing tensors route indices without changing the contracted scalar, making the network more suitable for boundary MPS operations.

Recent implementations have demonstrated the practical viability and close-to-optimal performance of this tensor-network-based decoding scheme~\cite{googlequantumai2023Suppressing}. Empirical results from state-of-the-art surface code experiments, including distance-5 logical qubits operating over 25 syndrome measurement rounds, reveal that the approximate likelihood converges rapidly. In practice, a relatively modest bond dimension of $\chi=30$ is sufficient to achieve convergence for the logical error probability. This highlights the efficiency of the MPS approximation in accurately resolving the dense, complex networks generated from realistic device noise.

\subsubsection{Beyond 2D: contraction strategies for 3D and circuit-level noise}
\label{sec:tn:3d}

The true challenge for TN decoders arises when addressing the three-dimensional structures produced by phenomenological noise, circuit-level noise, or 3D topological codes.
Contracting a 3D tensor network using standard MPS-based boundary methods introduces severe scaling challenges.
If the 3D network has dimensions $L \times L \times T$, contracting it slice by slice requires representing a 2D boundary using a Projected Entangled Pair State (PEPS).
Unfortunately, unlike 1D MPS, exact contraction of a 2D PEPS is itself \#P-hard, making the corresponding boundary compression operations algorithmically complex and computationally expensive.

Recent work by Piveteau, Chubb, and Renes~\cite{piveteau2023Tensorbeyond2d} overcame this barrier for the surface code by exploiting the structure of the 3D tensor network.
The key innovation was to recognize that the $L \times L \times T$ network has a favorable structure when contracted along one spatial direction rather than the temporal direction.
Contracting along the $x$-direction produces a boundary that is an $L \times T$ two-dimensional object.
Rather than representing this boundary as a PEPS (which would be intractable), they showed that it can be efficiently approximated as a matrix product operator (MPO) of length $T$, acting on an MPS of length $L$.
This ``MPO-MPS'' representation captures the essential temporal correlations while keeping the computational cost polynomial: $O(L \cdot T \cdot \chi^3)$ per contraction step.

They benchmarked this approach on the surface code with phenomenological noise and demonstrated near-optimal MLD approximation, significantly outperforming MWPM in the low-error regime and operating near the theoretical threshold.
At distance $d = 9$ and physical error rate $p = 2\%$, the TN decoder achieved logical error rates approximately $5\times$ lower than MWPM in their benchmarks, with the gap widening at larger distances~\cite{piveteau2023Tensorbeyond2d}.
However, extending this strategy to arbitrary circuit-level noise models with non-trivial spatial and temporal correlations remains a formidable challenge, primarily due to the intricate connectivity of the DEMs generated by realistic quantum circuits.

\subsubsection{Differentiable programming and noise estimation}
\label{sec:tn:differentiable}
Beyond serving as a powerful tool for inference, tensor networks can also be used for parameter estimation because the contraction is differentiable with respect to the local error probabilities. Cao \etal~\cite{cao2026differentiablemaximumlikelihoodnoise} use this observation in differentiable maximum-likelihood estimation (dMLE). Within the detector picture, the tensor network computes the marginal syndrome likelihood $\Prob(\mathbf{s})$ by summing over error mechanisms and logical variables rather than conditioning on a known true error. Because the local tensors depend smoothly on the physical noise parameters, automatic differentiation can propagate gradients through the contraction.

This full differentiability enables a purely data-driven optimization of the noise model directly from experimental logs. Given a dataset of experimentally observed syndromes $\mathcal{D} = \{\mathbf{s}\}$---without requiring any knowledge of the true underlying error configurations---we can formulate a differentiable loss function. Specifically, we optimize the noise parameters by minimizing the negative log-likelihood loss:
\begin{equation}
\mathcal{L}(\theta) = -\frac{1}{|\mathcal{D}|}\sum_{\mathbf{s} \in \mathcal{D}} \log P_\theta(\mathbf{s}),
\label{eq:dmle_loss}
\end{equation}
where $\theta$ denotes the physical parameters of the error DEM within the initial probability tensors.

By backpropagating the loss through the tensor network contraction, one can compute gradients of $\mathcal{L}(\theta)$ with respect to the initial tensor elements and update $\theta$ by gradient descent. The resulting parameters are maximum-likelihood estimates within the chosen DEM parameterization, providing hardware-tailored noise priors from syndrome data alone.

\subsection{Artificial intelligence perspective}
\label{sec:ai}

The third major approach to MLD seeks to approximate the optimal decoding distribution not by analytical mappings or structured tensor contractions, but by learning it directly from data.
By representing the decoding function as a deep neural network, the artificial intelligence (AI) perspective shifts the computational burden: a heavily intensive ``training'' phase is performed offline, producing a model that ideally enjoys fast, feed-forward inference online.

\subsubsection{Machine learning for QEC decoding: overview}
\label{sec:ai:overview}

Neural networks offer a powerful framework for approximating MLD. Instead of relying on explicit algorithmic search or tensor network contractions, neural decoders approach the task by parameterizing a conditional probability distribution $P_\theta(\boldsymbol{\ell} \mid \mathbf{s})$ over the logical sectors $\boldsymbol{\ell}$ given an observed syndrome $\mathbf{s}$.

When training such a network using a dataset sampled from the true physical error distribution, the standard objective is to optimize the network parameters $\theta$ by minimizing the cross-entropy loss:
\begin{equation}
\mathcal{L}(\theta) = -\mathbb{E}_{(\mathbf{s}, \boldsymbol{\ell}) \sim P_{\text{true}}} [\log P_\theta(\boldsymbol{\ell} \mid \mathbf{s})].
\end{equation}
Crucially, from an information-theoretic perspective, minimizing this cross-entropy loss is mathematically equivalent to minimizing the Kullback-Leibler (KL) divergence between the true posterior distribution $P_{\text{true}}(\boldsymbol{\ell} \mid \mathbf{s})$ and the network's parameterized distribution:
\begin{equation}
\arg\min_\theta \mathcal{L}(\theta) = \arg\min_\theta \mathbb{E}_{\mathbf{s}} \left[ D_{\text{KL}}\big(P_{\text{true}}(\boldsymbol{\ell} \mid \mathbf{s}) \parallel P_\theta(\boldsymbol{\ell} \mid \mathbf{s})\big) \right].
\end{equation}
By driving this KL divergence toward zero, the optimization process forces the neural network to align its internal representation with the true data distribution. In essence, the network implicitly learns to perform the computationally intractable summation over all degenerate error configurations within a logical sector. Consequently, the neural network effectively distills the complex, exponentially large MLD summation into an efficient, highly non-linear forward pass.

Furthermore, this data-driven approximation can adapt to hardware-specific noise correlations without manual redesign of the decoding logic.
The first wave of neural decoders included Boltzmann machine and probabilistic neural decoders~\cite{Torlai2017, KrastanovJiang2017}, feedforward and convolutional architectures for small surface-code instances~\cite{Varsamopoulos2017, Chamberland2020}, and methods targeted at correlated or device-specific noise~\cite{Baireuther2018}.
Subsequent work explored reinforcement-learning approaches, large-distance toric-code neural decoders, and scalable hierarchical architectures~\cite{Sweke2020, Ni2020, Meinerz2022}.
These early methods established the feasibility of learned decoding, but they also exposed the central scaling problem: a neural decoder must exploit code locality, time-translation symmetry, or logical-factorization structure to avoid becoming a small-distance classifier.

\subsubsection{Temporal translation symmetry and recurrent architectures}
\label{sec:ai:alphaqubit}
In modern QEC experiments, repeated syndrome measurements are essential to mitigate measurement errors that occur on ancilla qubits. Consequently, decoding under realistic circuit-level noise entails navigating a vastly expanded configuration space compared to idealized code-capacity models, for example, a transition from a 2D spatial lattice to a 3D spatio-temporal volume for the surface code. Fortunately, one does not necessarily have to process this complex, higher-dimensional decoding graph directly. By exploiting the inherent time-translation symmetry provided by the repeated measurement cycles, one can employ the recurrent structure of Recurrent Neural Networks (RNNs) to efficiently extract and process this temporal information. A remarkably successful approach to neural decoding was recently presented by Google Quantum AI and DeepMind with the AlphaQubit architecture~\cite{Bausch2024alphaqubit}.
Designed to tackle the full complexity of circuit-level noise and experimental data, AlphaQubit employs a three-component architecture:
\begin{enumerate}
    \item A linear ``tokenizer'' that embeds local syndrome measurements and supplementary experimental data into high-dimensional tokens.
    \item A spatial processing block that combines transformer-style attention with two-dimensional convolutional structure to capture correlations across the code patch.
    \item A recurrent update that carries hidden-state information across successive rounds of syndrome extraction.
\end{enumerate}

AlphaQubit is first trained on massive datasets generated from simulators using realistic circuit-level noise models.
Crucially, the model is then fine-tuned on actual experimental data from the Sycamore processor.
This allows the neural network to adapt to the specific crosstalk, leakage, and correlated error patterns of the physical hardware, which are often too complex to model accurately in standard simulators.

When tested on experimental data from Google's Sycamore processor, AlphaQubit outperformed state-of-the-art human-designed decoders on distance-3 and distance-5 surface-code data, including correlated matching, belief matching, and tensor-network baselines~\cite{Bausch2024alphaqubit, correlatedmatching2013, believematching2023}. Furthermore, as the focus of QEC expands toward more complex qLDPC codes, their highly connected, non-planar topologies present unique decoding challenges. Graph-based neural approaches such as GraphQEC~\cite{hu2025efficientuniversalneuralnetworkdecoder} directly use the stabilizer graph as input and have reported strong performance across surface, color, and qLDPC-code benchmarks. These results are promising, but for a review they should be treated as emerging evidence rather than settled replacements for conventional qLDPC decoders.

\subsubsection{Generative decoding for high-rate codes}
\label{sec:ai:generative}

Another formidable challenge in MLD emerges when decoding quantum codes with a large number of logical qubits ($k \gg 1$). For such codes, exact enumeration of all logical labels scales as $2^k$ for a single measured logical basis and as $4^k$ for full Pauli logical classes. Standard neural classifiers inherit this exponential output-space problem if they represent each logical sector as an independent class. A common approximation is to factorize the logical posterior into independent marginals, but this discards correlations between logical operators. To address this bottleneck, Cao \etal~\cite{cao2025generative} proposed Generative Neural Decoding (GND), which uses autoregressive generative modeling to represent the joint distribution of syndrome variables and a chosen set of binary logical labels:
\begin{equation}
P(\boldsymbol{\ell}, \mathbf{s}) = P(s_1)\prod_{i=2}^{m} P(s_i \mid s_{<i}) \prod_{j=1}^{k} P(\ell_j \mid \boldsymbol{s}, \ell_{<j}),
\label{eq:autoregressive}
\end{equation}
where $m$ is the length of syndrome.
Autoregressive neural networks, such as MADE~\cite{germain2015made} and Transformers~\cite{vaswani_attention_2017}, parameterize each conditional probability using a mask, ensuring that the prediction at position $t$ depends only on positions $1, \ldots, t-1$.
The training objective minimizes the negative log-likelihood over the dataset:
\begin{equation}
\mathcal{L}(\theta) = -\mathbb{E}_{E \sim \text{noise}}\big[\log P_\theta\big(\boldsymbol{\ell}(E),\, \mathbf{s}(E)\big)\big],
\label{eq:qecgpt_loss}
\end{equation}
where $\theta$ denotes the network parameters. During inference, the autoregressive model evaluates or generates logical strings sequentially, using
$P(\boldsymbol{\ell} \mid \mathbf{s}) = \prod_{i=1}^{k} P(\ell_i \mid \ell_{<i}, \mathbf{s})$.
This avoids a flat $2^k$-class output layer and can approximately perform MLD with a cost that scales linearly in the number of logical variables for the generation procedure reported in Ref.~\cite{cao2025generative}. The important caveat is that greedy sequential generation is an approximation to global maximum-a-posteriori inference unless augmented by search or sampling.

Recent empirical evaluations reported that GND improves over factorized neural baselines and conventional heuristic decoders on selected surface-code and qLDPC benchmarks~\cite{cao2025generative}. In a related but distinct direction, Blue \etal~\cite{blue2025machinelearningdecodingcircuitlevel} used a recurrent transformer-based decoder for circuit-level noise on bivariate bicycle (BB) codes, obtaining lower logical error rates than BP+OSD on the $[\![72,12,6]\!]$ BB code in the tested low-noise regime while maintaining a more predictable runtime.

\subsubsection{Scaling neural decoders to real-time}
\label{sec:ai:realtime}

Despite their accuracy, a major historical criticism of neural decoders has been their inference latency.
In a fault-tolerant setting based on superconducting qubits, syndrome extraction cycles occur every $\sim 1\,\mu$s.
To prevent an exponentially growing backlog of data (the so-called ``backlog problem''), the decoder must process syndromes at least as fast as they are generated.
This places an extremely tight computational budget, of order $10^6$ floating-point operations per syndrome round, on any decoder claiming real-time capability.

To meet the stringent real-time requirements of QEC, two distinct research trajectories have recently emerged. The first trajectory focuses on maximizing \textit{throughput} by leveraging the batch-processing capabilities of modern commercial accelerators like GPUs and TPUs. By decoding thousands of measurement rounds simultaneously, these methods achieve a highly competitive amortized decoding time per round. The second trajectory directly addresses strict \textit{latency} for closed-loop QEC, relying on highly customized hardware such as FPGAs and ASICs to deterministically generate a decoding result within the microsecond deadline. Both paradigms have recently yielded promising positive results.

In the throughput-oriented paradigm, recent works have optimized model architectures specifically for commercial accelerators. For instance, AlphaQubit 2~\cite{Bausch2024alphaqubit2} reported processing speeds faster than $1\,\mu$s per cycle for surface codes up to distance 11. It is important to note that this metric reflects an amortized throughput achieved via batching. While highly efficient for handling massive data streams, the physical overhead of transferring data to a GPU/TPU and accumulating a batch inherently differs from the microsecond-scale strict latency required for immediate real-time feedback. 

Similarly, parallelization through sliding-window decoding offers a robust method to handle continuous syndrome streams with high throughput. Zhang \etal~\cite{zhang2026learningdecodeparallelselfcoordinating} developed a self-coordinating neural network for parallel window decoding. By transforming the global decoding problem into a sequence of local, overlapping window predictions, they achieved highly parallelizable neural decoding. The network learns to encode its ``belief'' about the global error state into a compact hidden vector at each window boundary, which is passed to the adjacent window as additional context. While their reported TPU v6e benchmark shows sub-$1\,\mu$s/round speeds for simulated workloads up to $d=25$, this approach inherently remains a throughput-optimized batched pipeline.

Conversely, the latency-focused paradigm necessitates a shift from purely architectural enhancements to extreme hardware-algorithm co-design. A recent systematic study by Yan \etal~\cite{yan2026rethink} highlights that standard machine learning hardware is physically bottlenecked at the millisecond scale for unbatched inference. To address this, they focused on aggressive network pruning and quantization to generate ultra-lightweight neural decoders tailored for FPGAs. Crucially, this study demonstrates that extreme quantization to INT4 precision is essential to shift the computational burden from scarce digital signal processors to abundant look-up tables on FPGAs. Through this co-design approach, the authors illustrated that INT4-quantized and heavily pruned neural decoders can successfully survive the strict $1\,\mu$s latency deadline for surface codes up to distance $d=9$ on current FPGA hardware.

This latency problem has also motivated substantial work outside neural decoding.
Sparse Blossom and Fusion Blossom accelerate MWPM itself~\cite{higgott2023Sparse, wu2023fusion}, while modular, sliding-window, and parallel-window decoders restructure the decoding stream so that local windows can be processed in parallel with bounded latency~\cite{Bombin2023Modular, Skoric2023ParallelWindow, Tan2023SlidingWindow}.
Near-optimal ensembling methods such as harmonization provide another route: combine many fast approximate matching decoders, use agreement as a confidence measure, and reserve expensive decoding for ambiguous shots~\cite{Shutty2026Ensembling}.
These algorithmic developments are important because neural decoders must ultimately be compared not with textbook MWPM, but with the strongest real-time classical baselines.

\subsection{Synthesis of the three perspectives}
\label{sec:synthesis}

The three approaches above are best viewed as different computational realizations of the same coset-likelihood problem.
The statistical-mechanical formulation identifies the partition function and phase structure; tensor networks provide controlled contraction algorithms for that partition function; neural decoders learn an approximation to the posterior or to the logical decision rule from simulator and experimental data.
Table~\ref{tab:comparison} summarizes the practical distinctions that matter when choosing or evaluating an MLD or near-MLD method.
The most important distinction is what object the method actually computes or learns: an exact partition function, an approximate contraction, block-level beliefs, or a learned posterior/decision rule.

\begin{table}[t]
\centering
\footnotesize
\setlength{\tabcolsep}{2pt}
\caption{Comparison of representative approaches to MLD or near-MLD decoding.}
\label{tab:comparison}
\begin{tabular}{@{}L{0.18\linewidth} L{0.20\linewidth} L{0.18\linewidth} L{0.18\linewidth} L{0.20\linewidth}@{}}
\toprule
Approach & Object computed or learned & Best suited regime & Accuracy / runtime knob & Main bottleneck \\
\midrule
Planar exact solvers~\cite{PhysRevLett.134.190603} & Exact planar Ising partition functions $Z_{\boldsymbol{\ell}}$ & Repetition-code DEMs and planar code-capacity limits & Graph size and number of logical sectors & Requires planarity; not generic for circuit-level surface-code DEMs \\
2D TN contraction~\cite{Bravyi2014, Chubb2021tn2d} & Approximate coset partition functions & 2D local Pauli codes and code-capacity models & Bond dimension $\chi$ and contraction ordering & $\chi$ grows near threshold and with long-range correlations \\
3D TN contraction~\cite{piveteau2023Tensorbeyond2d} & Approximate spacetime likelihoods & Phenomenological or circuit-level surface-code data & Boundary ansatz, $\chi$, and window size & Boundary compression is harder than in 2D and often offline \\
BlockBP~\cite{KaufmannArad2025BlockBP} & Approximate coset beliefs on a block graph & 2D codes where local TN blocks are accurate & Block size and BP schedule & Loopy BP can fail under strong long-range correlations \\
Autoregressive neural decoders~\cite{Cao2023qecgpt, cao2025generative} & Learned joint distribution or posterior samples & High-rate or high-$k$ codes with large logical output spaces & Model size, ordering, search or sampling & Training coverage and approximation to the global optimum need validation \\
Recurrent transformer decoders~\cite{Bausch2024alphaqubit, Bausch2024alphaqubit2, blue2025machinelearningdecodingcircuitlevel} & Learned streaming logical decision rule & Repeated-measurement circuit-level data and real-time settings & Architecture, training data, windowing, accelerator & Generalization and certification outside training regimes are not automatic \\
\bottomrule
\end{tabular}
\end{table}
\section{Applications and outlook}
\label{sec:applications}

While optimal decoding is fundamentally a theoretical target, the development of scalable near-optimal decoders has already enabled applications beyond basic error correction, and a number of pressing challenges remain.

\subsection{Noise characterization and calibration}
\label{sec:apps:noise}
Conventionally, decoding and noise characterization are treated as separate tasks.
Hardware parameters are characterized using protocols like randomized benchmarking or gate set tomography, and the resulting error rates are fed into a decoder.
However, MLD-quality decoders invert this paradigm: the decoder itself becomes a powerful tool for characterizing the hardware.

The differentiable maximum-likelihood estimation (dMLE) framework~\cite{cao2026differentiablemaximumlikelihoodnoise} exemplifies this.
By maximizing the syndrome likelihood in Eq.~\eqref{eq:dmle_loss}, dMLE differentiates through exact or approximate likelihood evaluations to optimize circuit-level noise parameters.
Because this optimization targets the likelihood of the observed syndrome data, the learned parameters are decoder-independent within the assumed DEM parameterization and can subsequently be supplied to faster real-time decoders.

The practical impact of this approach is significant.
In the reported experiments, dMLE reduced logical error rates by up to about $30\%$ for repetition-code data and about $8\%$ for surface-code data by providing conventional decoders with better noise priors~\cite{cao2026differentiablemaximumlikelihoodnoise}.
This ``decode to characterize, characterize to decode better'' feedback loop, where the MLD-quality decoder extracts noise information that is then used to calibrate a faster heuristic decoder, represents a new paradigm for integrating decoding and hardware calibration.
Moreover, because the learned noise model captures qubit-specific and gate-specific error rates (including correlations), it provides hardware engineers with a detailed diagnostic map of the processor, highlighting which components contribute most to logical errors and guiding targeted hardware improvements.

\subsection{Experimental demonstrations}
\label{sec:apps:experiments}

The practical value of MLD and near-optimal decoders has been confirmed in several landmark quantum error correction experiments.
Early superconducting and trapped-ion experiments established repeated stabilizer measurement and real-time feedback in small codes~\cite{Kelly_2015, exp7_color_2021}, followed by repetition-code and surface-code scaling demonstrations on superconducting processors~\cite{googlerep, zhaoRealizationErrorCorrectingSurface2022, googlequantumai2023Suppressing}.
More recent platforms have expanded the experimental landscape to reconfigurable neutral-atom logical processors, trapped-ion logical circuits, and nonlocal-code demonstrations~\cite{bluvstein2023Logical, exp4_hyperproduct2024, reichardt2409demonstration, exp5_2024}.
Google Quantum AI's Willow experiment demonstrated below-threshold surface-code memories, including a distance-7 memory on a 105-qubit processor and a distance-5 memory with integrated real-time decoding on a 72-qubit processor~\cite{googlenew2024}.
The experiment also highlighted that offline, higher-accuracy decoders remain essential for benchmarking real-time decoders and diagnosing whether logical failures arise from hardware faults, model mismatch, or decoder suboptimality.

Subsequent numerical results on repetition codes further highlighted the importance of decoder choice. Cao \etal~\cite{PhysRevLett.134.190603} evaluated their exact polynomial Planar decoder on Google's repetition-code experiment dataset.
They revealed that a significant portion of the previously observed ``error floor'' was an artifact of the suboptimal heuristics in MWPM rather than an intrinsic limitation of the superconducting hardware itself.
This result had immediate practical implications: it showed that the physical qubits were performing better than the MWPM-based analysis had suggested, and that the path to lower logical error rates lay not only in hardware improvement but also in better classical post-processing.

More recently, neural decoders have moved from offline validation toward real-time feasibility.
AlphaQubit outperformed conventional decoders on Sycamore distance-3 and distance-5 surface-code data~\cite{Bausch2024alphaqubit}.
AlphaQubit 2 reported sub-microsecond-per-cycle throughput for selected surface- and color-code settings~\cite{Bausch2024alphaqubit2}.
A self-coordinating recurrent transformer reported real-time-scale parallel decoding on Zuchongzhi 3.2 surface-code data and TPU v6e simulations~\cite{zhang2026learningdecodeparallelselfcoordinating}.
Collectively, these experiments underscore that reaching fault-tolerance requires not only high-fidelity qubits but also decoding algorithms that fully capture the underlying noise correlations.

\subsection{Decoding beyond memory experiments}
\label{sec:apps:beyond}

To date, the most advanced MLD research has focused on QEC memory experiments, where the goal is to preserve a logical state over time.
However, universal fault-tolerant quantum computation requires executing logical gates (via lattice surgery, braiding, or transversal operations) and state distillation procedures.

These operations create geometrically complex, irregular spacetime decoding graphs that differ fundamentally from the regular lattice structures encountered in memory experiments.
For instance, lattice surgery involves dynamically merging and splitting code patches, which introduces new types of boundaries, time-varying code distances, and correlated measurement errors that traditional decoders struggle to handle optimally.
The merge and split operations create ``junction'' regions in the spacetime decoding graph where the local structure differs qualitatively from the bulk, and optimal decoding in these regions requires accounting for correlations that MWPM-based decoders cannot naturally represent.

The flexibility of neural network decoders makes them plausible candidates for these complex tasks, because they can in principle learn from irregular spacetime graphs rather than requiring hand-crafted matching rules.
This direction is beginning to move beyond speculation: the Multi-Core Circuit Decoder (MCCD) was recently proposed for logical circuits and reported competitive decoding accuracy with favorable runtime scaling on simulated random Clifford circuits containing single-qubit and entangling logical gates~\cite{Zhou2025LogicalCircuits}.
Nevertheless, scalable MLD-class decoding for full logical-gate workloads, especially with lattice-surgery layouts, state distillation, and hardware-calibrated circuit-level noise, remains an important open challenge.

Similarly, state distillation circuits, which are essential for implementing non-Clifford gates like the $T$ gate, produce complex, multi-level decoding problems where errors at different levels of the distillation hierarchy interact~\cite{BravyiKitaev2005, qec1fowlerintroduction2012}.
Developing MLD-class decoders that can handle logical-gate and distillation structures efficiently remains an important open challenge~\cite{qec2latticesurgery2012, BravyiKitaev2005}.

\subsection{Open challenges}
\label{sec:apps:challenges}

Despite the tremendous progress reviewed here, several key challenges remain for the field:

\compactheading{Scalability to large distances}
As hardware progresses to logical qubits protected by distance 15--31 codes (the regime required for practical algorithms), the computational cost of MLD approximations grows.
For TN decoders, the required bond dimension $\chi$ must increase to maintain accuracy on larger lattices, making $O(n\chi^3)$ scaling problematic at very large $n$.
For neural decoders, the quadratic scaling of self-attention with input sequence length becomes a bottleneck.
Hierarchical or multi-scale decoding approaches that partition the lattice into smaller, exactly solvable MLD modules, and combine their results using a coarser-grained global decoder, may be necessary to bridge the gap between near-optimal accuracy and practical computational budgets.

\compactheading{High-rate codes}
The transition from 2D topological codes to high-rate qLDPC codes~\cite{BreuckmannEberhardt2021} breaks the local geometric structure that both TN decoders and CNN-based neural decoders rely upon.
Recent asymptotically good qLDPC code families with $k = \Theta(n)$ and $d = \Theta(n)$ offer dramatically better encoding rates, but their expander-graph connectivity invalidates both planar-graph exact methods and local boundary-contraction strategies.
While generative decoding approaches (like GND) have shown early success on qLDPC codes~\cite{cao2025generative, blue2025machinelearningdecodingcircuitlevel}, developing scalable, degeneracy-aware decoders for arbitrary expander-graph codes remains a critical open problem.
New ideas from the theory of approximate tensor network contraction on expander graphs, or graph neural network architectures that respect the code's algebraic structure, may be needed.

\compactheading{Hardware co-design}
Decoding algorithms should not be developed in isolation from hardware constraints.
The integration of specialized decoding units (such as neuromorphic chips, dedicated ASICs, or FPGAs) directly into the cryogenic control loop of the quantum processor will require MLD algorithms to be heavily quantized, pruned, and optimized for low-latency integer arithmetic.
The ``software-defined decoder'' paradigm, where a trained neural network is compiled into a fixed-function hardware pipeline, is a promising direction, but requires careful co-optimization of the network architecture, the training procedure, and the target hardware platform.

\compactheading{Correlated and non-Pauli noise}
Most MLD formulations assume that the noise channel is a Pauli channel (or can be approximated as one via Pauli twirling).
Real hardware exhibits non-Pauli noise mechanisms including coherent errors (systematic over-rotations), leakage (population escaping the computational subspace), and crosstalk (unwanted couplings between qubits that are not being actively gated).
Extending MLD to handle these noise types, either by generalizing the partition function formulation or by training neural decoders on data that includes non-Pauli effects, is an important direction for bridging the gap between theoretical decoding and practical hardware operation.

\section{Conclusion}
\label{sec:conclusion}

MLD serves as the gold standard for quantum error correction.
While computing the exact MLD probability is formally \#P-hard for general codes, the realization that MLD is not practically intractable for the codes and noise models of greatest interest has sparked a renaissance in decoder design.
This topical review has explored three synergistic approaches that approximate the same underlying MLD objective through different computational lenses:

\begin{enumerate}
  \item \textbf{Statistical mechanics} provides the theoretical foundation, mapping QEC to classical disordered spin models. This mapping identifies error thresholds with phase transitions along the Nishimori line and, remarkably, yields exact, polynomial-time MLD solutions for planar scenarios such as the repetition code under circuit-level noise~\cite{PhysRevLett.134.190603}.
  \item \textbf{Tensor networks} provide the algorithmic framework, casting the evaluation of coset probabilities as the contraction of a factor graph. Advanced boundary-contraction techniques derived from condensed matter physics permit near-optimal decoding of 2D and 3D code structures, with accuracy systematically controlled by the bond dimension $\chi$~\cite{Bravyi2014, Chubb2021tn2d, piveteau2023Tensorbeyond2d}.
  \item \textbf{Artificial intelligence} provides a scalable implementation route. By shifting much of the computational burden to an offline training phase, generative models~\cite{cao2025generative} and recurrent transformers~\cite{Bausch2024alphaqubit, blue2025machinelearningdecodingcircuitlevel} learn complex error distributions directly from data, achieving state-of-the-art accuracy on several real-hardware and realistic-simulation benchmarks and moving toward real-time operability~\cite{Bausch2024alphaqubit2, zhang2026learningdecodeparallelselfcoordinating}.
\end{enumerate}

Rather than competing, these three paradigms are increasingly converging.
Differentiable tensor networks are trained using machine-learning techniques to estimate noise parameters~\cite{cao2026differentiablemaximumlikelihoodnoise}; neural networks use simulator and tensor-network data for pre-training~\cite{Bausch2024alphaqubit}; and statistical-mechanical factor graphs provide the block-belief-propagation structures used to accelerate inference~\cite{KaufmannArad2025BlockBP}.
As quantum processors scale from the current era of early logical qubits ($d = 3$--$7$) into the fault-tolerant regime ($d = 15$--$31$ and beyond), this unified approach to MLD will be essential to extracting the maximum possible logical fidelity from noisy physical qubits, and ultimately to realizing the transformative promise of quantum computation.

\section*{Acknowledgments}
This work is supported by the Ministry of Education Singapore under grant No. SKI 2021\_07\_03, National Quantum Computing Hub translational fund No. W24Q3D0002, MTC Young Individual Research Grant No. H25-MRG3466, and Advanced Quantum Algorithm and Solutions Grant No. S25Q9DA001.

\bibliographystyle{unsrtnat}
\bibliography{ref}

\end{document}